\documentclass[aps,prd,showpacs,floatfix,notitlepage,nofootinbib,twocolumn,a4paper]{revtex4}
\usepackage{amsfonts,amsmath,units,wasysym,epsfig,graphicx,verbatim,color,subfigure,graphicx,bm,mathrsfs,lipsum,hyperref,cleveref}
\usepackage{booktabs}
\usepackage{orcidlink}
\usepackage[normalem]{ulem}  
\definecolor{dgreen}{RGB}{0, 204, 0}

\newcommand{\be}{\begin{equation}}
\newcommand{\ee}{\end{equation}}
\newcommand{\bea}{\begin{eqnarray}}
\newcommand{\eea}{\end{eqnarray}}

\usepackage{multirow}

\DeclareUnicodeCharacter{2212}{-}
\usepackage[none]{hyphenat}

\usepackage{cellspace}
\begin{document}

\title{Sensitivity of Neutron Star Observables to Transition Density in Hybrid Equation-of-State Models}


\author{N. K. Patra\orcidlink{0000-0003-0103-5590}$^{1}$}
\email{nareshkumarpatra@cuhk.edu.cn}

\author{Sk Md Adil Imam\orcidlink{0000-0003-3308-2615}$^{2}$}
\email{adil.imam@unab.cl}

\author{Kai Zhou\orcidlink{0000-0001-9859-1758}$^{1,3}$}
\email{zhoukai@cuhk.edu.cn }

\affiliation{$^1$School of Science and Engineering, The Chinese University of Hong Kong, Shenzhen, Guangdong, 518172, China}

\affiliation{$^2$Instituto de Astrof\'isica, Departamento de F\'isica y Astronom\'ia, Universidad Andr\'es Bello,Santiago, Chile}

\affiliation{$^3$School of Artificial Intelligence, The Chinese University of Hong Kong, Shenzhen, Guangdong, 518172, China}

\date{\today}

\begin{abstract} 
We investigate how the transition density (\(\rho_{\rm tr}\)) affects hybrid constructions of the neutron star (NS) equation of state (EoS)
in which a nucleonic description at low densities is matched to a model-agnostic high-density extension based on a speed-of-sound (CS) parametrization. Using four representative nucleonic models--Taylor expansion, \(\frac{n}{3}\) expansion, Skyrme, and relativistic mean-field--built from identical nuclear matter parameters, we isolate the impact of the low-density EoS and the transition density on NS observables. We find that, within the present smooth-matching prescription, NS properties such as radii and tidal deformabilities retain significant sensitivity to the choice of low-density EoS for commonly adopted transition densities around \(\rho_{\rm tr} \approx 2\rho_0\), even when the same high-density parametrization is employed. This residual dependence arises from differences in the matching conditions at \(\rho_{\rm tr}\), which propagate into the high-density extension, so different low-density inputs lead to different effective high-density EoSs. These findings are robust across two qualitatively distinct CS parametrizations differing in peak height, location, and width. Quantitatively, the model spread in radius and tidal deformability at $1.4\,M_\odot$ exceeds the current observational uncertainty {($\sigma_{R_{1.4}}$ = 0.8 km and $\sigma_{\Lambda_{1.4}}$= 200)} by factors of $\sim 1.8$ and $\sim 1.4$ at $\rho_{\rm tr} \approx 2\rho_0$, whereas these factors reduce to $\sim 1.05$ and $\sim 0.4$ at $\rho_{\rm tr} = \rho_0$. Lowering the transition density, therefore, systematically diminishes the spread among models and leads to more consistent predictions. Our results demonstrate that the widely used choice \(\rho_{\rm tr} \approx 2\rho_0\) does not guarantee practical model independence within the present hybrid EoS framework, and should be treated as an explicit source of systematic uncertainty when inferring dense matter properties from NS observations. 
\end{abstract}

\maketitle

\section{Introduction} \label{introduction}

Neutron stars (NSs) are the only compact objects in the Universe that serve as natural laboratories to study ultra-dense, highly isospin-asymmetric matter \cite{Glendenning:1997wn, Lattimer:2006xb, Haensel2007, Oertel:2016bki, Baym:2017whm, Rezzolla:2018jee}. The equation of state (EoS) of dense matter plays a central role in this context, directly determining their macroscopic properties and linking nuclear physics to astrophysical observations \cite{Lattimer:2012nd, Hebeler:2013nza, Lattimer:2021emm, Ferreira:2021pni, Zhang:2018vrx, OmanaKuttan:2022aml}. While nucleonic degrees of freedom dominate near nuclear saturation density ($\rho_0 \simeq 0.16~\text{fm}^{-3}$), matter in NS cores is governed by $\beta$-equilibrium and charge neutrality \cite{Li:2019tcx, Alford:2018lhf, Khunjua:2020hbd}, and at suprasaturation densities ($\rho > \rho_0$) may transition to exotic phases such as hyperons, meson condensates, or deconfined quark matter \cite{Zhao:2020dvu, Malik:2022jqc, Albino:2024ymc, Bonanno:2011ch, Benic:2014jia, Raduta:2022elz, Miyatsu:2015kwa, Li:2025obt}.

Multimessenger observations have significantly constrained the EoS in recent years \cite{Abbott:2017vwq, Abbott:2018exr, Abbot2019}. Gravitational wave detections from binary NS mergers (GW170817, GW190425) have measured tidal deformability, while NICER has provided precise mass-radius measurements for pulsars including PSR J0437+4715 \cite{Choudhury:2024xbk}, PSR J0614+3329 \cite{Mauviard:2025dmd}, PSR J0030+0451 \cite{Riley:2019yda, Miller:2019cac}, and PSR J0740+6620 \cite{Riley:2021pdl, Miller:2021qha}. The observation of massive NSs with $M \gtrsim 2\,M_\odot$ further demands a stiff high-density EoS. Future detectors, including the Einstein Telescope \cite{Punturo:2010zz} and Cosmic Explorer \cite{Reitze:2019iox}, are expected to sharpen these constraints considerably. Despite these advances, the behavior of matter at suprasaturation densities---particularly the possible appearance of exotic degrees of freedom---remains poorly constrained, motivating continued study of hybrid star EoS models incorporating hadronic-to-quark or hadronic-to-hyperonic transitions \cite{Most:2018eaw, Han:2018mtj, Komoltsev:2024lcr, Huang:2025vfl, Xia:2024wpz, Verma:2025vkk, Zhou:2024yzy, Celi:2025zmn, Li:2025obt}.

Such transitions are typically modeled as first-order phase transitions or smooth crossovers, with the transition density often assumed to lie in the range $\sim 2$--$4\rho_0$. However, conclusions drawn from these studies frequently conflict, reflecting their strong dependence on modeling assumptions, choice of interactions, and imposed observational constraints.A systematic framework that minimizes model dependence is therefore essential. Speed-of-sound (CS) parametrizations 
\cite{Tews:2018kmu, Greif:2018njt, Tews:2018iwm} provide a flexible model-agnostic framework for extending the EoS from nucleonic descriptions to higher densities while preserving thermodynamic stability and causality,
and naturally connect to perturbative QCD predictions ($c_s^2 \to 1/3$) at asymptotically high densities \cite{Kurkela:2009gj, Annala:2019puf, Annala:2023cwx, Fujimoto:2022ohj, Sasaki:2011ff}. The requirement of supporting 
$M_{\rm max} \geq 2\,M_\odot$ constrains the amount of softening permitted at high density and motivates exploring smooth CS extensions, but it does not generically exclude strong first-order phase transitions if sufficiently rapid restiffening occurs above the transition.

In this work, we investigate the dependence of neutron star observables on both the low-density EoS and the transition density. The transition density ($\rho_{\rm tr}$) is defined as the matching point between a low-density nucleonic EoS and a high-density CS parametrization \cite{Tews:2018kmu}. Below $\rho_{\rm tr}$, we employ four nucleonic models---Taylor \cite{Chen:2005ti, Chen:2009wv, Newton:2014iha, Margueron:2017eqc, Margueron:2018eob}, $n/3$ \cite{Imam:2021dbe, Lattimer:2015nhk, Gil:2016ryz}, Skyrme \cite{Dutra:2012mb, Chabanat:1997qh, Margueron:2002fq, Gomez:1992dgj, Agrawal:2005ix}, and relativistic mean-field (RMF) \cite{Shen:1998gq, Dutra:2014qga, Walecka:1974qa}---constructed with an identical set of nuclear matter parameters (NMPs), isolating the effect of the functional form. We systematically vary the transition density as $\rho_{\rm tr} = \rho_0,\, 1.5\rho_0,\, 2\rho_0$ and quantify the resulting impact on NS properties.{Throughout this work, $\rho_{\rm tr}=\rho_0$ is employed as a limiting reference point for studying the dependence of neutron-star observables on the matching density. Although such an early transition is not intended to represent a realistic physical scenario, it provides a useful benchmark because the nucleonic EoS is most reliably constrained in the vicinity of saturation density.}

While CS parametrizations have been widely adopted in Bayesian analyses \cite{Capano:2019eae, Dietrich:2020efo, Venneti:2026lpp, Imam:2025lut}, the role of $\rho_{\rm tr}$ as an independent source of systematic uncertainty has not been systematically isolated in a controlled framework; its effect is entangled with simultaneous variations in NMPs, CS coefficients, and observational inputs. {Ref.~\cite{Venneti:2026lpp} investigated hybrid EoSs within a
Bayesian framework by matching an RMF-based
low-density EoS to the high-density CS
parametrization, while varying the transition density
$\rho_{\rm tr}$ together with the remaining EoS
parameters. Consequently, the impact of $\rho_{\rm tr}$
was intertwined with variations in the underlying EoS
parameters and observational constraints. In contrast,
the present work adopts a controlled, non-Bayesian
approach in which the nuclear matter parameters are
held fixed and four distinct nucleonic functional forms
(Taylor, $n/3$-expansion, Skyrme, and RMF) are
matched to the same high-density CS parametrization.
By systematically varying only $\rho_{\rm tr}$, we
directly isolate the residual dependence of neutron-star
observables arising from the matching procedure itself.} {Although the present work employs the same four
nucleonic EoSs (Taylor, n/3, Skyrme, and RMF) and
the same CS parametrization used in Ref.~\cite{Imam:2025lut},
their roles are different. In Ref.~\cite{Imam:2025lut}, the
nucleonic EoSs were treated as complete EoSs which started after the crust ($0.5\rho_0$) and
extrapolated to the stellar center, while the CS construction was considered as a separate model above the fixed transition density, 2$\rho_0$. In contrast, here each nucleonic EoS is employed after the crust
only up to a transition density $\rho_{\rm tr}$ and is then matched to a common high-density CS parametrization.} The present work therefore addresses this gap through a controlled, non-Bayesian study: by matching four qualitatively distinct nucleonic models with identical NMPs to the same high-density CS parametrization at systematically varied $\rho_{\rm tr}$, we directly quantify how much residual model dependence in NS observables originates from the matching procedure itself. Our results reveal that this dependence is substantial and has been systematically underestimated in prior hybrid EoS constructions.

The paper is organized as follows. In Section~\ref{formalism}, we outline the formalism. Section~\ref{results} presents the results and discussion, and Section~\ref{Summary} provides a brief summary. Throughout the paper, we consider $G=c=\hbar=1$.

\section{Methodology} \label{formalism}

We construct the EoS in two density regimes separated by the transition density $\rho_{\rm tr}$. Below $\rho_{\rm tr}$, four nucleonic models are employed — Taylor, $n/3$-expansion, Skyrme, and RMF — following the methodology detailed in Ref.~\cite{Imam:2025lut}. Beyond $\rho_{\rm tr}$, the EoS is extended using a speed-of-sound parametrization that ensures thermodynamic stability, causality, and the approach to the conformal limit $c_s^2 \to 1/3$ at asymptotically large densities.

\subsection{Equation of State at Low and High Densities}

Following Ref.~\cite{Tews:2018kmu}, the speed of sound for $\rho > \rho_{\rm tr}$ is parametrized as,
\bea
c_s^2 &=& \frac{1}{3} - c_1 \exp\left[{-\frac{(\rho-c_2)^2}{n_b^2}}\right] + h_p \exp\left[-\frac{(\rho-n_p)^2}{w_p^2}\right]\nonumber\\
&& \times\left[1 + \mathrm{erf}\!\left(s_p \frac{\rho-n_p}{w_p}\right)\right], \label{eq-vs}
\eea
{The parameters $h_p$, $n_p$, $w_p$, $s_p$, and $n_b$
are treated as independent inputs that control the
height, location, width, skewness, and asymptotic
behavior of the speed-of-sound profile, respectively.
For a given choice of these parameters and a specified
transition density $\rho_{\rm tr}$, the coefficients
$c_1$ and $c_2$ are not independent. Instead, they are
uniquely determined by requiring continuity of
$c_s^2(\rho)$ and its first derivative at the matching
density,$\rho_{\rm tr}$. Consequently, only
$\{h_p,n_p,w_p,s_p,n_b\}$ constitute independent CS
parameters, while $c_1$ and $c_2$ are matching
coefficients fixed by the low-density EoS and the
continuity conditions. The adopted nuclear matter
parameters (NMPs) together with the independent CS
parameters are listed in Table~\ref{tab1}, whereas the resulting
values of $c_1$ and $c_2$ for each low-density EoS and transition density are reported in Tables~\ref{tab2} and \ref{tab3}.}

Starting from $\rho_{\rm tr}$, where the energy density $\varepsilon(\rho_{\rm tr})$, pressure $P(\rho_{\rm tr})$, and derivative $\varepsilon'(\rho_{\rm tr})$ are specified from the nucleonic EoS, the high-density EoS is built iteratively with step size $\Delta\rho = 0.001~\mathrm{fm}^{-3}$:
\bea
\rho_{i+1} &=& \rho_i + \Delta\rho, \label{eq-rhoE}\\
\varepsilon_{i+1} &=& \varepsilon_i + \Delta\rho\,\frac{\varepsilon_i + P_i}{\rho_i}, \label{eq-engE}\\
P_{i+1} &=& P_i + c_s^2(\rho_i)\,\Delta\varepsilon, \label{eq-preE}
\eea
where $i=0$ corresponds to $\rho_{\rm tr}$, and $\Delta\varepsilon$ is evaluated using the zero-temperature thermodynamic relation $P = \rho\,(\partial\varepsilon/\partial\rho) - \varepsilon$.

\begin{table}[b]
\caption{\label{tab1} The nuclear matter parameters (in MeV) and CS parametrization coefficients. The saturation density, $\rho_0$ and $n_b$, $n_p$, $w_p$ are in fm$^{-3}$; $h_p$ and $s_p$ are dimensionless. {Note that the coefficients $c_1$ and
$c_2$ are not free parameters; for a given low-density
EoS and transition density $\rho_{\rm tr}$, they are
uniquely determined by imposing continuity of $c_s^2$
and its first derivative at $\rho_{\rm tr}$.}}
\centering
\setlength{\tabcolsep}{3.0pt}
\renewcommand{\arraystretch}{1.4}
\begin{ruledtabular}
\begin{tabular}{ccccccccc}
 & {$\rho_0$} & {$e_0$} & {$K_0$} & {$Q_0$} & {$J_0$} & {$L_0$} & {$K_{\rm sym,0}$} \\[1.3ex]
\hline
Set1 & 0.159 & -15.58 & 264.13 & -368.34 & 31.12 & 43.86 & -71.69 \\[1.3ex]
Set2 &  &  & Same & as  & Set1  &  &   \\[1.3ex]
\hline\hline
 & {$n_b$} & {$h_p$} & {$n_p$} & {$w_p$} & {$s_p$} & & \\[1.3ex]
\hline
Set1 & 0.705 & 0.53 & 3.99 & 4.66 & -29.87 & & \\[1.3ex]
Set2 & 0.277 & 0.66 & 2.48 & 1.41 & -42.96 & & \\[1.3ex]
\end{tabular}
\end{ruledtabular}
\end{table}

\section{Results and Discussion}\label{results}
\begin{figure}
    \centering
    \includegraphics[width=0.9\linewidth]{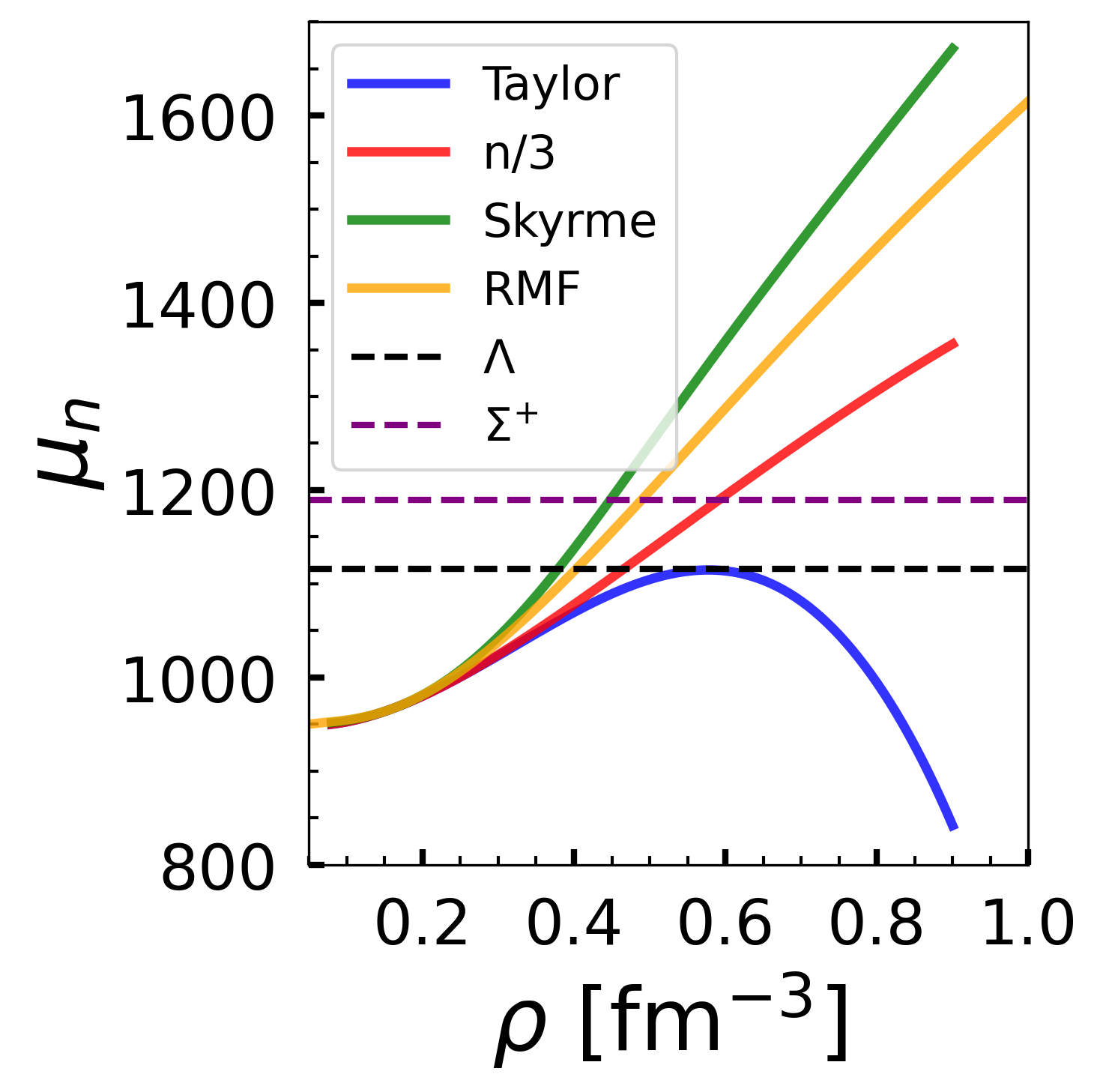}
    \caption{
Baryon chemical potential as a function of density for the four nucleonic equations of state (Taylor, $n/3$, Skyrme, and RMF) constructed with identical nuclear matter parameters. While the models are consistent around saturation density $\rho_0$, significant deviations appear at higher densities, reflecting the model dependence of the extrapolation beyond $\rho_0$. These differences at the matching point influence the construction of the high-density EoS and the resulting neutron star observables.
}
    \label{fig1}
\end{figure}

\begin{figure*}
    \centering
    \includegraphics[width=\linewidth]{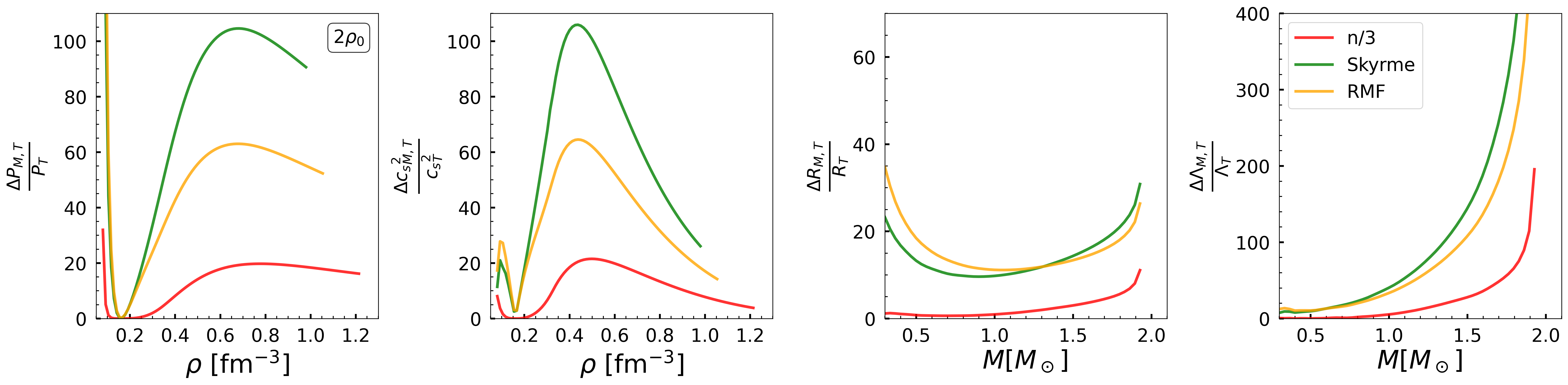}
    \includegraphics[width=\linewidth]{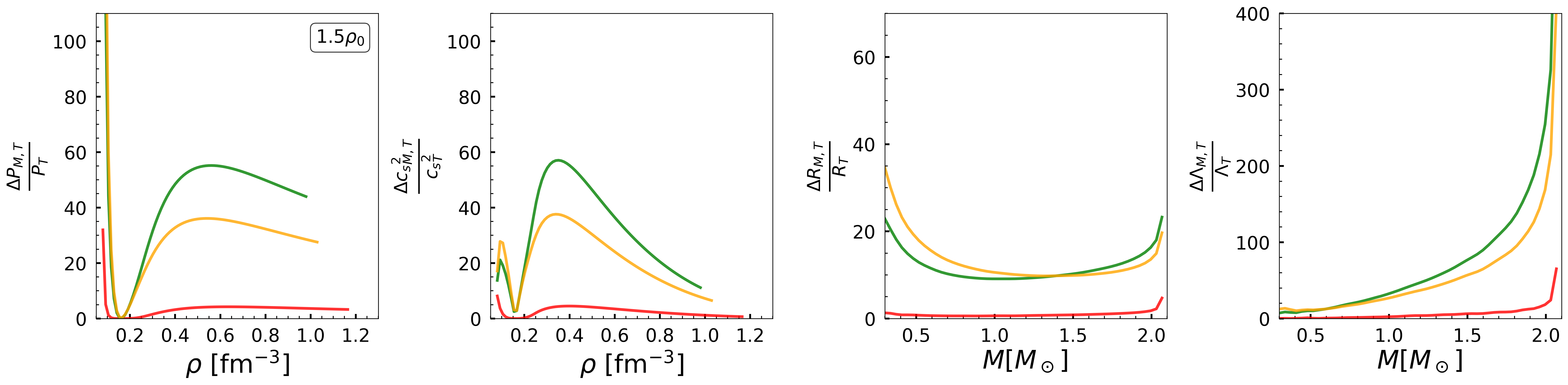}
    \includegraphics[width=\linewidth]{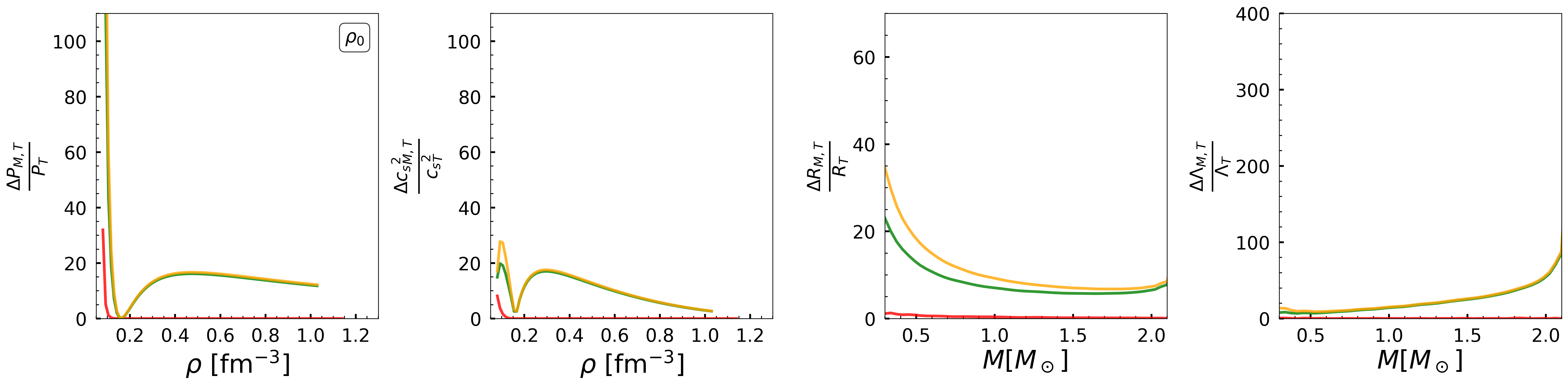}
    \caption{
Equation of state and neutron star observables for different transition densities $\rho_{\rm tr} = 2\rho_0$, $1.5\rho_0$, and $\rho_0$ (top to bottom). Results are shown for four nucleonic models (Taylor, $n/3$, Skyrme, and RMF) matched to the same speed-of-sound (CS) parametrization above $\rho_{\rm tr}$. From left to right: relative deviations in pressure, squared speed of sound as a function of density, relative deviations in radius and tidal deformability as a function of mass. The relative deviations are computed with respect to the Taylor+CS baseline. Significant model dependence persists at higher transition densities, particularly in radii and tidal deformabilities, while lower $\rho_{\rm tr}$ reduces these differences.
}
    \label{fig2}
\end{figure*}

\begin{figure*}
    \centering
    \includegraphics[width=\linewidth]{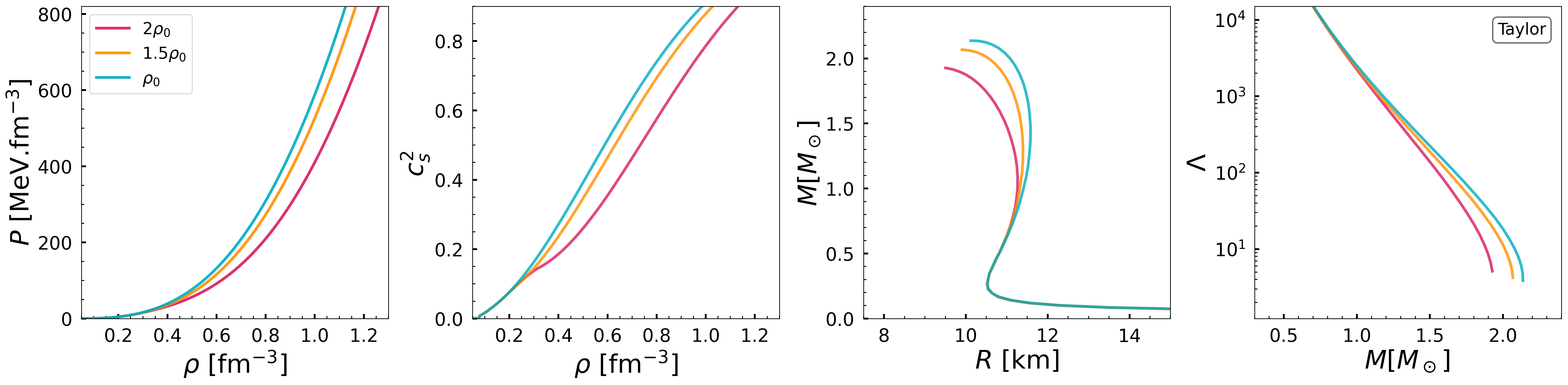}
    \caption{
Equation of state and neutron star observables for a fixed low-density nucleonic model matched to the same speed-of-sound parametrization at different transition densities $\rho_{\rm tr} = 2\rho_0$, $1.5\rho_0$, and $\rho_0$. From left to right: pressure as a function of density, squared speed of sound, mass--radius relation, and tidal deformability as a function of mass. Even with identical low- and high-density inputs, varying $\rho_{\rm tr}$ leads to significant differences in the EoS and neutron star properties. Lower transition densities result in a stiffer high-density EoS, yielding larger radii, higher maximum masses, and increased tidal deformabilities.
}
    \label{fig3}
\end{figure*}
{We consider four nucleonic EoS models---Taylor,
$n/3$-expansion, Skyrme, and RMF---constructed as in
Ref.~\cite{Imam:2025lut}, combined with the Baym, Pethick, and Sutherland
model~\cite{Baym:1971pw} for the outer crust and a
polytropic form~\cite{Carriere:2002bx} for the inner
crust. The core EoS is built from these models up to
$\rho_{\rm tr}$ and matched to the high-density CS EoS.
This differs from the framework of
Ref.~\cite{Imam:2025lut}, where the nucleonic EoSs
were treated as complete EoSs and extrapolated to the
stellar center. Here, all four nucleonic models are
matched to the same high-density CS parametrization
at $\rho_{\rm tr}$, enabling a direct assessment of the
residual dependence on the low-density EoS.} Imposing thermodynamic stability, causality ($c_s^2 < 1$), and positive symmetry energy ( up to $\rho_{\rm tr}$ for the nucleonic EoS) ensures physically consistent EoSs, which are then employed to compute neutron star properties (mass, radius, and tidal deformability) by solving the Tolman--Oppenheimer--Volkoff (TOV) equations~\cite{Oppenheimer:1939ne, Tolman:1939jz}.

Massive pulsars with $M \geq 2M_\odot$~\cite{Miller:2021qha, Riley:2021pdl, Romani:2022jhd, Linares:2018ppq} imply central densities of $\sim 5$--$6\rho_0$, well beyond the regime where nucleonic descriptions are reliable. As shown in Fig.~\ref{fig1}, the neutron chemical potential exceeds the $\Lambda$ hyperon mass at $\sim 2$--$3\rho_0$ (model-dependent), consistent with expectations for hyperon onset~\cite{Weissenborn:2011ut, Malik:2022jqc, Chatterjee:2015pua}. The actual threshold in $\beta-$equilibrated matter also depends on in-medium interactions, charge chemical potentials, and many-body effects, so this comparison is qualitative. This motivates model-agnostic approaches such as transitioning to a CS parametrization near $2\rho_0$~\cite{Tews:2018kmu, Tews:2018iwm, Capano:2019eae, Dietrich:2020efo, Venneti:2026lpp,Reed:2025sqh}.

In this work, we test whether $\rho_{\rm tr} = 2\rho_0$ renders NS observables insensitive to the low-density EoS, and if not, assess whether there exists a range of transition densities for which the residual sensitivity of neutron star observables to the low-density EoS becomes subdominant relative to the adopted observational uncertainty scale.{While transition densities in the range $2$--$4\rho_0$
are commonly assumed in hybrid-EoS studies, the
present work is intentionally restricted to
$\rho_{\rm tr}\le2\rho_0$. Our primary objective is to
investigate how uncertainties associated with the
low-density nucleonic EoS propagate into neutron-star
observables through the matching procedure. At
densities significantly above $2\rho_0$, additional
degrees of freedom such as hyperons or other exotic
components may become relevant, rendering a purely
nucleonic description increasingly uncertain.
Consequently, we focus on the density regime where
the nucleonic sector can be examined in a controlled
manner.
}

We use the four nucleonic models with identical NMPs (Table~\ref{tab1}). Results for Set1 are shown in the main text; Set2 results are in the Appendix. The two sets represent qualitatively distinct high-density behaviors: Set1 features a broad, low-amplitude CS peak at high density ($n_p \approx 3.99$ fm$^{-3}$), while Set2 has a sharper, higher-amplitude peak at lower density ($n_p \approx 2.48$ fm$^{-3}$), sampling two different regimes of the high-density CS landscape. We note that the EoSs are considered 
up to 1.4 fm$^{-3}$, which is above the central density of the maximum mass NS. The consistency of conclusions across both sets (see Appendix) indicates that the observed transition density dependence is not an artifact of a particular CS parametrization. A full Bayesian exploration of the CS parameter space is deferred to future work.

Figure~\ref{fig2} shows the EoS and NS observables as a function of transition density. To quantify model dependence, we define the relative deviation of model $M$ with respect to the Taylor+CS baseline:
\[
\frac{\Delta X_{M,T}}{X_T} = \frac{abs(X_{M+\mathrm{CS}} - X_{T+\mathrm{CS}})}{X_{T+\mathrm{CS}}} \times 100\%.
\]
Even with identical saturation properties, differences in the crust and at the transition point lead to different matching coefficients $(c_1, c_2)$, producing distinct high-density EoSs.

At $\rho_{\rm tr} = 2\rho_0$, the $n/3$ model shows the smallest deviation ($\lesssim 20\%$), while RMF and Skyrme exhibit significantly larger radii. RMF dominates below $\rho_0$ and Skyrme above, with a crossing near $1.3\,M_\odot$ (central density $\sim 3\rho_0$), confirming that radii are primarily governed by the EoS {around and
moderately above } saturation density~\cite{Alam:2016cli, Patra:2022yqc, Patra:2023jvv, Patra:2023jbz, Imam:2023ngm, Kunjipurayil:2022zah, Reed:2021nqk, Malik:2020vwo, Tsang:2020lmb}. Tidal deformability deviations reach $\sim 400\%$ at high masses, demonstrating that NS observables can differ substantially across low-density EoS models even with identical NMPs and CS parametrization. {Since all four nucleonic EoSs are calibrated to the
same nuclear matter parameters, they agree most
closely in the vicinity of saturation density and diverge
progressively as their functional forms are extrapolated
to higher densities.}

Reducing $\rho_{\rm tr}$ to $1.5\rho_0$ decreases the maximum pressure deviation from $\sim 100\%$ to $\sim 60\%$. The Taylor and $n/3$ models yield nearly identical radii, while Skyrme and RMF still show larger values, though with reduced deviations at high masses. At $\rho_{\rm tr} = \rho_0$, Taylor and $n/3$ fully coincide above $\rho_0$, and pressure deviations for Skyrme and RMF fall to $\lesssim 20\%$, with radii closely approaching the Taylor results. In all cases, for low-mass star ($\sim 0.3\,M_\odot$) the radius differences remain unchanged, since the sub-$\rho_0$ EoS is unaffected by the choice of $\rho_{\rm tr}$.

To quantify when NS observables become effectively model-independent, we define
\begin{equation}
\mathcal{R}_{X_{1.4}}(\rho_{\rm tr}) = \frac{\max X_{1.4} - \min X_{1.4}}{\sigma_{X_{1.4}}},
\end{equation}
where the numerator is the maximum spread across low-density EoS models at $1.4\,M_\odot$ and $\sigma_{X_{1.4}}$ is the current $68\%$ credible interval (CI) observational uncertainty. For $R_{1.4}$ we compute the average error,$\sigma_{ R_{1.4}}=0.8$ km from $R_{1.4} = 11.36^{+0.95}_{-0.63}~\mathrm{km}$~\cite{Choudhury:2024xbk}, and for $\Lambda_{1.4}$ we convert the $90\%$ CI result $\Lambda_{1.4} = 190^{+390}_{-120}$~\cite{Abbott:2018exr} to a $68\%$ CI, yielding $\sigma_{\Lambda_{1.4}} \approx 200$\footnote{The posterior of Ref.~\cite{Abbott:2018exr} is non-Gaussian; the conversion to $68\%$ CI is an approximation. However, varying $\sigma_{\Lambda_{1.4}}$ within a reasonable range does not alter the qualitative trend of $\mathcal{R}_{\Lambda_{1.4}}$ decreasing with lower $\rho_{\rm tr}$, and the conclusions are robust.}. We define \textit{practical model independence} as $\mathcal{R}_{X_{1.4}} < 1$, i.e., the spread across models falls below current observational uncertainty. When $\mathcal{R} > 1$, model dependence is resolvable by current observations and must be accounted for in inference analyses.

\begin{table*}[]
\centering
\caption{The variation of neutron star properties with transition densities ($\rho_{\rm tr}$) and low density EoS models for Set1. Here $c_1, c_2$(in fm$^{-3}$) (Eq.\ref{eq-vs}) are required to have continuity of speed of sound and its derivative at the transition density. $\mathcal{R}_{X_{1.4}}(\rho_{\rm tr}) = \frac{\max X_{1.4} - \min X_{1.4}}{\sigma_{X_{1.4}}}$,  X being the NS observable (radius or tidal deformability) and $\sigma_{X_{1.4}}$ is the average uncertainty at 68\% credible interval for the observable X at $1.4 M_\odot$ NS. These uncertainties are taken to be $\sigma_{R_{1.4}} = 0.8,\mathrm{km}$ and $\sigma_{\Lambda_{1.4}} = 200$ . To assess the sensitivity of our results to these choices, we vary both uncertainties by $\pm 25\%$(for R$_{1.4}$ : $\pm$0.2 km, for $\Lambda_{1.4}: \pm$50), referring to the resulting cases as the low and high uncertainty scenarios (see the text for details).}

\label{tab2}
\renewcommand{\arraystretch}{1.4}
\begin{ruledtabular}
\begin{tabular}{c c c c c c c c c ccc ccc}
$\rho_{\rm tr}$ & Models & $c_1$ & $c_2$ &
$M_{\max}$ & $R_{1.4}$&
$\Lambda_{1.4}$ &
$\rho_c(1.4)$ &
$\rho_c({\max})$ &
\multicolumn{3}{c}{$\mathcal{R}_{R_{1.4}}(\rho_{\rm tr})$} &
\multicolumn{3}{c}{$\mathcal{R}_{\Lambda_{1.4}}(\rho_{\rm tr})$} \\

\cline{10-15}

&&&&&&&&&
$\sigma_{R_{1.4},low}$ &
$\sigma_{R_{1.4}}$ &
$\sigma_{R_{1.4},high}$ &
$\sigma_{\Lambda_{1.4},low}$ &
$\sigma_{\Lambda_{1.4}}$ &
$\sigma_{\Lambda_{1.4},high}$ \\

\hline

\multirow{4}{*}{$2\rho_0$}
& Taylor  & 0.7669 & 0.2581 & 1.92 & 11.08 & 240.20 & 0.62 & 1.27
& \multirow{4}{*}{2.39}
& \multirow{4}{*}{1.79}
& \multirow{4}{*}{1.43}
& \multirow{4}{*}{1.81}
& \multirow{4}{*}{1.36}
& \multirow{4}{*}{1.08} \\
& $n/3$   & 0.7850 & 0.1641 & 2.02 & 11.36 & 293.52 & 0.55 & 1.22
& & & & & & \\
& Skyrme  & 1.0887 & -0.1889 & 2.33 & 12.52 & 512.54 & 0.40 & 0.98
& & & & & & \\
& RMF     & 0.8272 & 0.0195 & 2.19 & 12.47 & 448.58 & 0.44 & 1.06
& & & & & & \\

\hline

\multirow{4}{*}{$1.5\rho_0$}
& Taylor  & 0.8211 & 0.1001 & 2.06 & 11.38 & 305.31 & 0.53 & 1.18
& \multirow{4}{*}{1.53}
& \multirow{4}{*}{1.40}
& \multirow{4}{*}{0.91}
& \multirow{4}{*}{1.32}
& \multirow{4}{*}{0.99}
& \multirow{4}{*}{0.79} \\
& $n/3$   & 0.8301 & 0.0787 & 2.08 & 11.47 & 321.23 & 0.52 & 1.17
& & & & & & \\
& Skyrme  & 1.0267 & -0.1540 & 2.31 & 12.50 & 503.82 & 0.40 & 0.99
& & & & & & \\
& RMF     & 0.9023 & -0.0509 & 2.23 & 12.49 & 455.30 & 0.43 & 1.03
& & & & & & \\

\hline

\multirow{4}{*}{$\rho_0$}
& Taylor  & 0.8552 & 0.0307 & 2.13 & 11.58 & 353.08 & 0.49 & 1.14
& \multirow{4}{*}{1.40}
& \multirow{4}{*}{1.05}
& \multirow{4}{*}{0.83}
& \multirow{4}{*}{0.56}
& \multirow{4}{*}{0.42}
& \multirow{4}{*}{0.34} \\
& $n/3$   & 0.8553 & 0.0307 & 2.13 & 11.60 & 350.94 & 0.49 & 1.14
& & & & & & \\
& Skyrme  & 0.8942 & -0.0374 & 2.21 & 12.12 & 434.97 & 0.44 & 1.04
& & & & & & \\
& RMF     & 0.8959 & -0.0397 & 2.21 & 12.42 & 435.94 & 0.44 & 1.04
& & & & & & \\

\end{tabular}
\end{ruledtabular}
\end{table*}

From Table~\ref{tab2}, at $\rho_{\rm tr} = 2\rho_0$, both $\mathcal{R}_{R_{1.4}} = 1.79$ and $\mathcal{R}_{\Lambda_{1.4}} = 1.36$ exceed unity, indicating significant resolvable model dependence. At $\rho_{\rm tr} = 1.5\rho_0$, the values reduce to $\mathcal{R}_{R_{1.4}} = 1.40$ and $\mathcal{R}_{\Lambda_{1.4}} = 0.99$, with tidal deformability marginally crossing the threshold. At $\rho_{\rm tr} = \rho_0$, $\mathcal{R}_{\Lambda_{1.4}} = 0.42 < 1$ achieves practical model independence, while $\mathcal{R}_{R_{1.4}} = 1.05$ marginally exceeds unity. {These results are for $\sigma_{ R_{1.4}}=0.8$ km  and $\sigma_{\Lambda_{1.4}}$ =200.} { To test the sensitivity of the practical-independence
criterion to the adopted observational uncertainties, we
varied both $\sigma_{R_{1.4}}$ and
$\sigma_{\Lambda_{1.4}}$ by $\pm25\%$. The resulting
values of $\mathcal{R}_{R_{1.4}}$ and
$\mathcal{R}_{\Lambda_{1.4}}$ are listed in Tables~\ref{tab2} and
\ref{tab3}. While the precise numerical values change, the
qualitative trend remains unchanged: lower transition
densities consistently yield smaller intermodel spreads
in NS observables. We therefore conclude that $\rho_{\rm tr}=\rho_0$
corresponds to a near minimum of the residual
dependence on the selected low-density EoSs within
the adopted hybrid-EoS framework. These results confirm that commonly used transition
densities around $2\rho_0$ do not eliminate the
residual dependence of NS observables on the
selected low-density EoSs. The observed convergence
at lower transition densities should therefore be
interpreted as a reduction of this residual dependence
within the adopted hybrid-EoS framework, rather than
as a reduction of the overall uncertainty associated
with dense matter.
We note that all constructed EoSs are retained when
evaluating the spread measures reported in this work,
including cases with $M_{\rm max}<2\,M_\odot$. The
objective is to quantify the residual dependence of
NS observables on the matching procedure
and the choice of low-density EoS, rather than to
construct an ensemble prefiltered by observational
constraints.
}

Figure~\ref{fig3} isolates the effect of $\rho_{\rm tr}$ by fixing the low-density model (Taylor) and CS parametrization while varying $\rho_{\rm tr} = \rho_0,~1.5\rho_0,~2\rho_0$. Lower transition densities lead to an earlier onset of the stiffer CS EoS, producing larger radii, higher maximum masses, and larger tidal deformabilities, while higher $\rho_{\rm tr}$ retains the softer nucleonic behavior to larger densities, yielding more compact configurations.

The physical origin lies in the matching procedure. For fixed CS parameters $(h_p, n_p, w_p, s_p, n_b)$, continuity of $c_s^2$ and its derivative at $\rho_{\rm tr}$ uniquely determines $c_1$ and $c_2$. Since the nucleonic EoS yields different values of $P(\rho_{\rm tr})$, $\varepsilon(\rho_{\rm tr})$, and $c_s^2(\rho_{\rm tr})$ at different transition densities, the coefficients $c_1$ and $c_2$ — and hence the effective high-density EoS — differ across choices of $\rho_{\rm tr}$ (Table~\ref{tab2}). At lower $\rho_{\rm tr}$, the handoff occurs at lower pressure and energy density, yielding values of $c_1$ and $c_2$ that produce a stiffer effective EoS, propagating into larger radii, higher maximum masses, and increased tidal deformabilities.

This highlights a key subtlety often overlooked in hybrid EoS constructions: the CS parametrization does not uniquely determine the high-density EoS — it does so only in conjunction with a specific $\rho_{\rm tr}$ through the matching-induced $c_1$ and $c_2$. Consequently, constraints on CS parameters inferred from NS observations are implicitly conditioned on the assumed transition density. This underscores the need to treat $\rho_{\rm tr}$ as a free parameter in future Bayesian analyses.

\section{Conclusions and Outlook}\label{Summary}
We have examined the dependence of NS observables on the transition density $\rho_{\rm tr}$ in hybrid EoS constructions, where a nucleonic EoS is matched to a high-density CS parametrization. By employing four low-density models with identical NMPs, we isolate the roles of $\rho_{\rm tr}$ and the matching procedure as independent sources of systematic uncertainty.

Our key finding is that, within the present smooth-matching framework and for the commonly adopted representative parameter sets considered here, NS radii and tidal deformabilities remain significantly sensitive to the low-density EoS at $\rho_{\rm tr} \approx 2\rho_0$, even when the high-density parametrization is identical. This sensitivity originates in the matching conditions, which differ across models and produce distinct high-density extensions despite identical CS parameters. Lowering $\rho_{\rm tr}$ reduces these discrepancies and yields more consistent NS predictions. We note that $\rho_{\rm tr} = \rho_0$ should be regarded as a mathematical limiting case rather than a physical prescription, since the nucleonic EoS is most reliably constrained near saturation density. Transition densities in the range $\rho_0 \lesssim \rho_{\rm tr} \lesssim 1.5\rho_0$ therefore offer a better balance between {practical} model independence and physical reliability than the commonly adopted $\rho_{\rm tr} \approx 2\rho_0$.{This behavior is expected because the four nucleonic
EoSs share the same nuclear matter parameters and
therefore agree most closely near saturation density,
while differences emerge primarily from their
high-density extrapolations.} The robustness of these findings across two qualitatively distinct CS parameter sets — differing in peak height, location, and width — suggests that this behavior is not tied to a specific CS parametrization and is likely a generic feature of the matching procedure.

These results have direct implications for model-agnostic Bayesian analyses~\cite{Capano:2019eae, Dietrich:2020efo, Venneti:2026lpp}, which typically fix or marginalize over $\rho_{\rm tr}$ without quantifying the associated systematic bias. Our study demonstrates that $\rho_{\rm tr}$ is not a benign modeling choice but a source of uncertainty that can exceed current observational precision, and should ideally be varied and marginalized over rather than fixed a priori in future inference analyses that simultaneously vary the transition density, EoS parametrizations, and observational inputs from gravitational wave and X-ray measurements. A natural extension is to identify NS observables that are insensitive to the low-density EoS and map them directly onto CS parameters, facilitating efficient Bayesian inference

\section{Author Contributions}
NKP and AI conceived the project, carried out the calculations, and prepared the initial draft; both contributed equally as first authors. KZ contributed to the discussion, as well as to the writing and revision of the manuscript.

\section{Acknowledgements}
KZ acknowledges support by the NSFC grant under No. 92570117, the CUHK-Shenzhen University development fund under grant No. UDF01003041 and UDF03003041, Shenzhen Peacock fund under No. 2023TC0179 and the Ministry of Science and Technology of China under Grant No. 2024YFA1611004.

\section{Conflicts of Interest} The authors declare no conflict of interest.
%


\section*{\Large Appendix: Results for Set2} \label{appendix}

In this Appendix, we repeat the analysis presented in the main text {using the same set of nuclear matter parameters  but different CS parameters(Set2),} as listed in Tab\ref{tab1}. The purpose is to assess the robustness of our conclusions against variations in the CS parameters input.

The qualitative behavior remains unchanged. In particular, neutron star observables continue to exhibit a significant dependence on the choice of low-density EoS when the transition density is taken to be $\rho_{\rm tr} \approx 2\rho_0$, despite employing the same high-density speed-of-sound parametrization. As in the main results, this dependence originates from differences in the matching conditions at the transition density.

\begin{figure*}
    \centering
    \includegraphics[width=\linewidth]{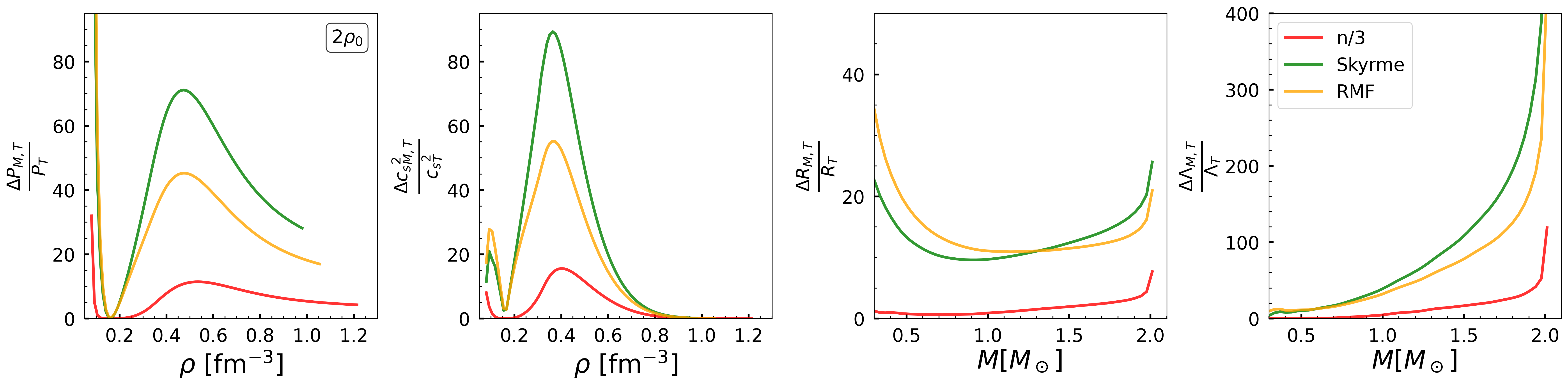}
    \includegraphics[width=\linewidth]{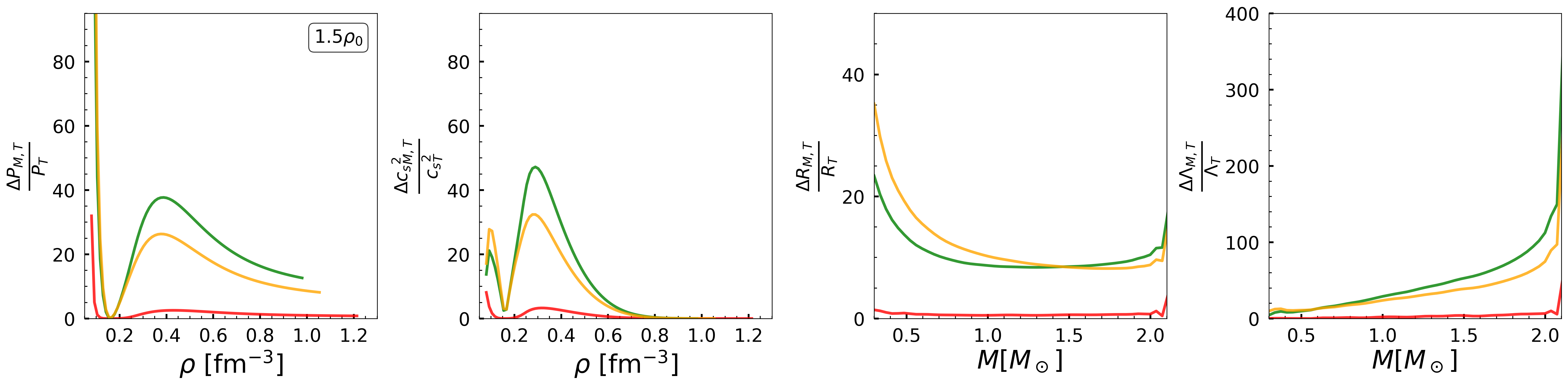}
    \includegraphics[width=\linewidth]{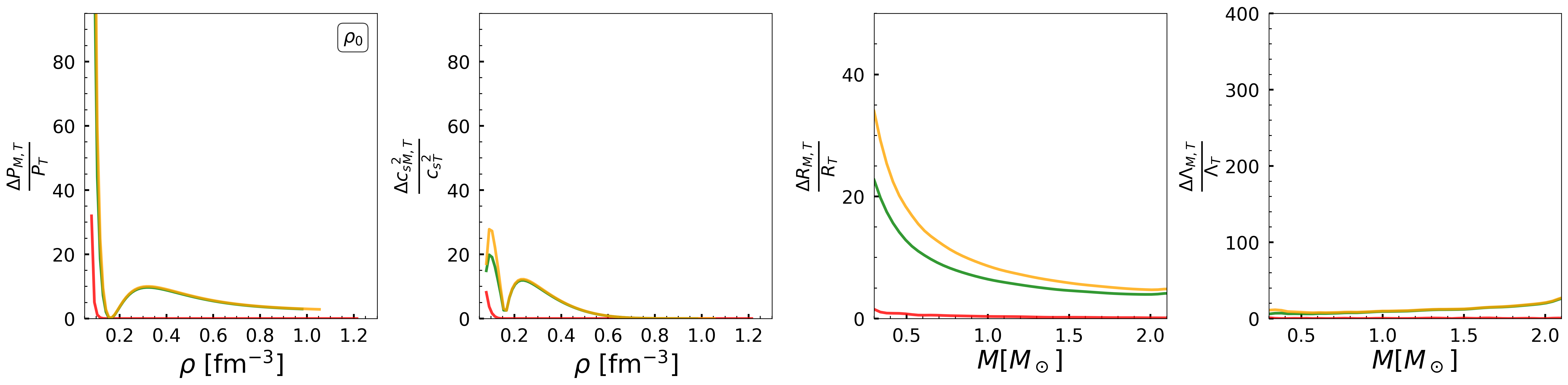}
    \caption{Same as Fig. \ref{fig2}, but, for results using Set2 parameters as listed in Table~\ref{tab1}.}
    \label{fig4}
\end{figure*}

\begin{figure*}
    \centering
    \includegraphics[width=\linewidth]{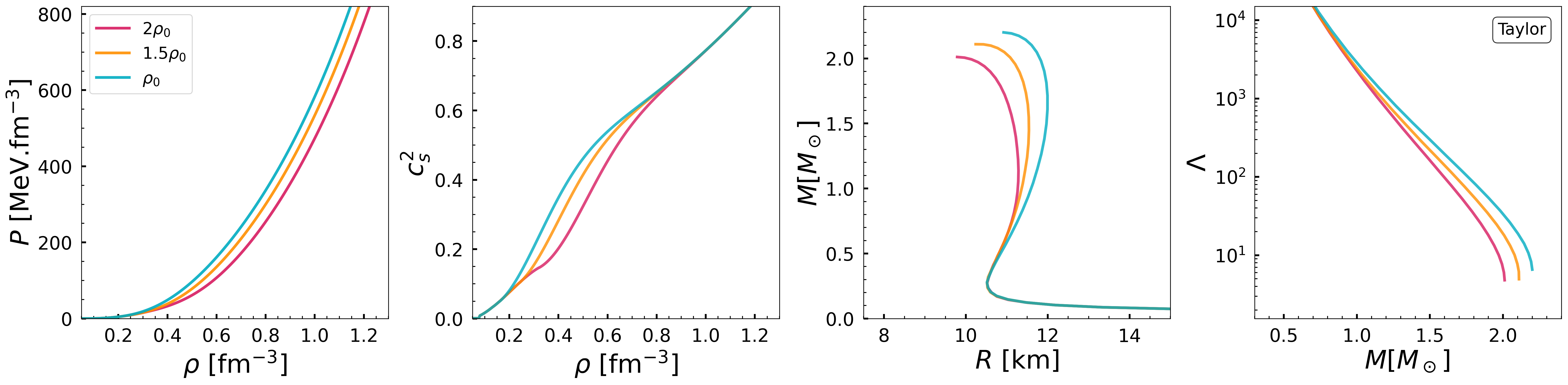}
    \caption{Same as Fig. \ref{fig3}, but, for results using Set2 parameters as listed in Table~\ref{tab1}.}
    \label{fig5}
\end{figure*}

Reducing the transition density systematically decreases the spread in observables across different low-density models, leading to more consistent predictions. This trend is clearly reflected in Tab~\ref{tab3} and Figs.~\ref{fig4} and~\ref{fig5}, and is fully consistent with the findings obtained for Set1.

{These results confirm that our main conclusion---namely, that commonly used transition densities around
$2\rho_0$ do not eliminate the residual dependence of neutron-star observables on the selected low-density
EoSs within the adopted hybrid-EoS framework---is robust with respect to variations in CS parametrization.}

\begin{table*}[t]
\centering
\caption{Same as Table \ref{tab2}, but, for results using Set2 parameters as listed in Table~\ref{tab1}.}
\label{tab3}
\centering    
\setlength{\tabcolsep}{3.0pt}
\renewcommand{\arraystretch}{1.4}
\begin{ruledtabular}
\begin{tabular}{c c c c c c c c c ccc ccc}
$\rho_{\rm tr}$ & Models & $c_1$ & $c_2$ &
$M_{\max}$ & $R_{1.4}$&
$\Lambda_{1.4}$ &
$\rho_c(1.4)$ &
$\rho_c({\max})$ &
\multicolumn{3}{c}{$\mathcal{R}_{R_{1.4}}(\rho_{\rm tr})$} &
\multicolumn{3}{c}{$\mathcal{R}_{\Lambda_{1.4}}(\rho_{\rm tr})$} \\

\cline{10-15}

&&&&&&&&&
$\sigma_{R_{1.4},low}$ &
$\sigma_{R_{1.4}}$ &
$\sigma_{R_{1.4},high}$ &
$\sigma_{\Lambda_{1.4},low}$ &
$\sigma_{\Lambda_{1.4}}$ &
$\sigma_{\Lambda_{1.4},high}$ \\

\hline

\multirow{4}{*}{$2\rho_0$}
& Taylor  & 0.3151 & 0.3068 & 2.01 & 11.21 & 267.99 & 0.57 & 1.22 & \multirow{4}{*}{2.18}
& \multirow{4}{*}{1.64}
& \multirow{4}{*}{1.30}
& \multirow{4}{*}{1.62}
& \multirow{4}{*}{1.22}
& \multirow{4}{*}{0.97} \\
& $n/3$   & 0.3106 & 0.2707 & 2.05 & 11.41 & 306.46 & 0.53 & 1.14 &  &  \\
& Skyrme  & 0.4183 & 0.0824 & 2.23 & 12.52 & 512.27 & 0.40 & 0.97 &  & \\
& RMF     & 0.2922 & 0.2016 & 2.15 & 12.47 & 449.71 & 0.44 & 1.06 &  & \\

\hline  

\multirow{4}{*}{$1.5\rho_0$}
& Taylor  & 0.3467 & 0.1949 & 2.10 & 11.53 & 343.95 & 0.49 & 1.14 &  \multirow{4}{*}{1.66}
& \multirow{4}{*}{1.25}
& \multirow{4}{*}{1.0}
& \multirow{4}{*}{1.08}
& \multirow{4}{*}{0.81}
& \multirow{4}{*}{0.65} \\
& $n/3$   & 0.3486 & 0.1870 & 2.11 & 11.60 & 355.70 & 0.48 & 1.06 &  & \\
& Skyrme  & 0.3963 & 0.0938 & 2.23 & 12.50 & 506.75 & 0.40 & 0.97 &  & \\
& RMF     & 0.3568 & 0.1354 & 2.19 & 12.53 & 466.16 & 0.42 & 0.98 &  & \\

\hline  

\multirow{4}{*}{$\rho_0$}
& Taylor  & 0.3818 & 0.1173 & 2.20 & 11.92 & 453.50 & 0.42 & 0.97 &  \multirow{4}{*}{1.23}
& \multirow{4}{*}{0.93}
& \multirow{4}{*}{0.74}
& \multirow{4}{*}{0.37}
& \multirow{4}{*}{0.28}
& \multirow{4}{*}{0.22} \\
& $n/3$   & 0.3818 & 0.1173 & 2.20 & 11.95 & 454.72 & 0.42 & 0.97 &  & \\
& Skyrme  & 0.3945 & 0.0940 & 2.23 & 12.50 & 506.39 & 0.40 & 0.97 &  & \\
& RMF     & 0.3950 & 0.0932 & 2.23 & 12.66 & 509.15 & 0.40 & 0.97 &  & \\

\end{tabular}
\end{ruledtabular}
\end{table*}

\end{document}